\documentclass{elsart3p}
\usepackage{graphicx}
\usepackage{amssymb}

\begin{document}

\begin{frontmatter}

\title{Direct Photon Production in p+p and d+Au Collisions Measured with the PHENIX Experiment}

\author{D. Peressounko for the PHENIX collaboration\thanksref{authList}}
\thanks[authList]{For the full list of PHENIX authors and acknowledgements, see \cite{PHENIX-AuAu}.}
\address{RRC "Kurchatov Institute", Kurchatov sq. 1, 123182, Moscow, Russia\thanksref{ead}}
\thanks[ead]{{\it Email address:} peressou@rcf.rhic.bnl.gov}

\begin{abstract}

Direct photon yield measured in p+p and d+Au collisions at $\sqrt{s} = 200$~GeV with the PHENIX experiment is presented and compared to pQCD predictions. In the case of p+p collisions both the inclusive and isolated direct photon production cross-sections are measured. Relative contribution of isolated direct photons is somewhat below but still agree within errors with pQCD predictions. In the case of d+Au collisions no modifications due to presence of cold matter are found within errors. 

\end{abstract}

\end{frontmatter}

Direct photons weakly interact with matter and being emitted almost surely escape from hot zone without further interactions. This makes them an ideal probe of the initial state of hadron+hadron or A+A collision and a good tool for test of pQCD predictions.

Presently a somewhat controversial situation appeared with theoretical description of the experimental data on direct photon production in $p+p$ and $p+\bar p$ collisions. One hand, NLO calculations with joint recoil and threshold resummation describe well most of the existing data on direct photon production in p+p and $p+\bar p$ collisions \cite{pqcd-comp}. On the other hand, there are several datasets (mostly fixed target low energy experiments \cite{R806,E706}) which considerably deviate from the theoretical predictions and require additional $k_T$ kick to reproduce the data. In this context the new PHENIX data are important since they cover intermediate energy range between existing experiments.

In the leading order in $\alpha_s$ direct photon emission in p+p collisions is described by Compton-like ($q+g\to q+\gamma$) and annihilation ($q+\bar q \to g+\gamma$) processes. In addition to these "direct" processes photons can be emitted in parton fragmentation, which is formally next order in $\alpha_s$, but gains amplification due to collinear singularity (note that higher order corrections smooth out difference between two types of processes). In contrast to photons, emitted in "direct" processes, photons originated in fragmentation processes will be accompanied by hadrons. Therefore one can experimentally estimate relative contribution of "direct" and "fragmentation" processes measuring inclusive and isolated direct photon yield. To our knowledge PHENIX is the first experiment measured both inclusive and isolated direct photon yields and provided possibility to compare these two contributions.

The advantages of direct photons are revealed most clearly in p+A and A+A collisions, where they provide possibility to fix initial state and to test inner region of the collision. PHENIX has measured direct photon yield in p+p and d+Au collisions and thus we are able to look at modifications of structure functions in nuclei. Moreover, these data can be used as a baseline for A+A collisions and provide possibility to extract matter-related part of the photon radiation, such as thermal emission or jet-matter interaction.

Detailed description of the PHENIX experiment can be found in \cite{PHENIX-NIM}. Photons are measured by electromagnetic calorimeter EMCal, consisting of 8 sectors: 2 sectors of lead glass (PbGl) and 6 sectors of lead scintillator (PbSc). EMCal covers $2\times 90^\circ$ in azimuthal angle and $|\eta|<0.35$ in pseudorapidity. Neutral pions and $\eta$ mesons are measured through their $2\gamma$ decays with the same calorimeter. Registration of charged particles and measurement of their momenta are performed by PHENIX tracking system, including drift chamber (DC) and 3 layers of Pad Chambers (PC). Both DC and PC cover same azimuthal and pseudorapidity range as EMCal. Event triggering and centrality determination in the case of d+Au collisions are done with Beam-Beam Counter (BBC) and Zero Degree Calorimeter (ZDC). 

\begin{figure}
\includegraphics[width=\columnwidth]{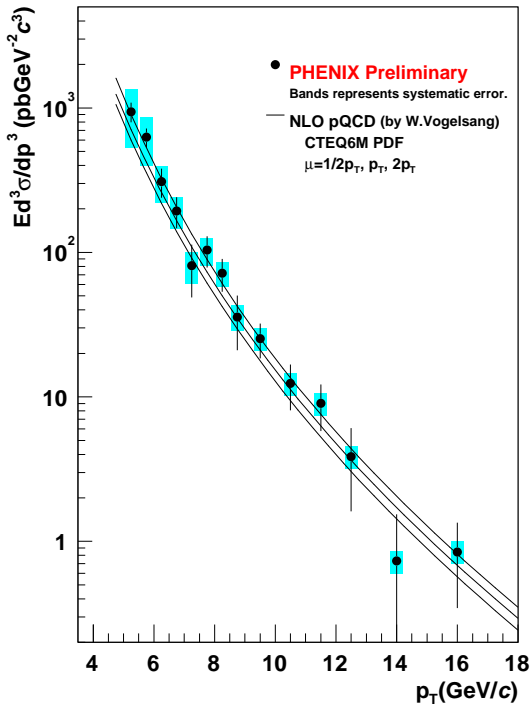} 
\caption{Inclusive direct photon production cross-section, measured in p+p collisions at $\sqrt{s}=200$ GeV compared to 
NLO pQCD predictions \cite{Vogel} calculated at different scales.}
\label{fig:pp-subtr}
\end{figure}

Inclusive direct photon yield is measured by subtracting decay photons from the total photon yield. The total photon yield is given by measurement of photon-like showers in EMCal, correcting for efficiency and contamination due to charged and neutral hadrons. The decay photon spectrum is calculated from measured $\pi^0$ and $\eta$ spectra and includes as well minor contribution of heavier hadron decays. In a low multiplicity environment of p+p and d+Au collisions we improve signal/background ratio by applying tagging method. Within this method, in each event we exclude photons, which have $\pi^0$ decay photons in the detector and then subtract remaining contribution of decay photons in a way similar to subtraction method. We proved that both methods provide consisting results.

Inclusive direct photons production cross-section in p+p collisions at $\sqrt{s}=200$~GeV is presented in fig.~\ref{fig:pp-subtr}. It is based on an integrated luminosity of 266 nb$^{-1}$ collected in RHIC Run-3. We compare our data to NLO pQCD predictions \cite{Vogel} including recoil and threshold resummations and using CTEQ6M parton distribution factions and GRV parameterization of fragmentation functions. Calculations are made for 3 scales of factorization-fragmentation-renormalization scales: $\mu_{f}=\mu_{fr}=\mu_r=\mu=1/2p_T,\, p_T,\, 2p_T$. In the full range of measurements $5\le p_T \le 16$ GeV we find good agreement with pQCD predictions so that no additional $k_T$ kick is required to reproduce our data.    

\begin{figure}
\includegraphics[angle=-90,width=\columnwidth]{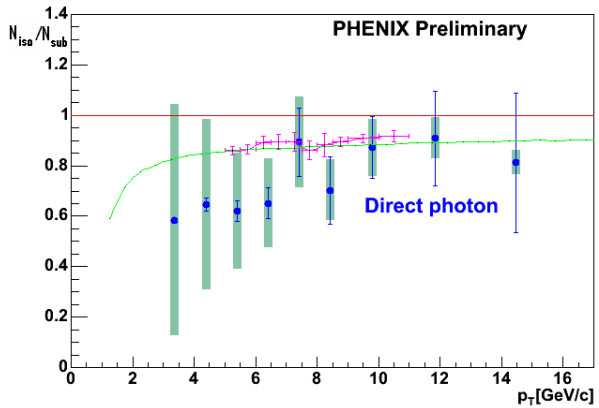} 
\caption{Ratio of isolated to inclusive direct photon yield, measured in p+p collisions at $\sqrt{s}=200$ GeV compared to 
pQCD predictions of V.Vogelsang \cite{Vogel} (line) and M.Werlen \cite{Werlen} (crosses).}
\label{fig:iso-sub}
\end{figure}

One can gain more insight into the details of direct photon production in p+p collisions by comparing inclusive and isolated photon yield. Isolation cut reduces contribution of fragmentation processes leaving mostly photons, emitted in "direct" process. However, result depends on the particular isolation cut used in the analysis. Most of the previous analyses apply the following isolation cut: total energy in the cone of some radius around the photon $R=\sqrt{\Delta \phi^2+\Delta \eta^2}$ (where $\Delta\phi$ is the azimuthal angle difference and $\Delta\eta$ is the pseudorapidity difference) should be below some fraction ($E_{cone}/E_\gamma\le 0.1$) of the photon energy. Typically radius $R=0.4$ [rad] was used. Due to the PHENIX limited acceptance we define the following isolation cone:
$$
R=\sqrt{\Delta\eta^2 + \Delta\phi^2} \le 0.5;\, |\eta|<0.35. 
$$
That is we increased cone radius to compensate our limited acceptance in $\eta$ by larger coverage in $\phi$. We impose usual cut on total energy in the cone $E_{cone}<0.1\cdot E_\gamma$. To estimate $E_{cone}$ we add momenta of charged hadrons measured in DC and total energy of neutral particles measured in calorimeter providing that these clusters do not match with any track in DC. Once we have a list of isolated photons we then remove decay photons in the way, similar to those, used in tagging method. We present ratio of isolated to inclusive direct photon yields in fig.~\ref{fig:iso-sub} and compare it to two pQCD predictions. The first one \cite{Werlen} is done using JETPHOX Monte-Carlo program \cite{JETPHOX} with CTEQ6M structure functions, BFG set2 fragmentation functions and all scales set to $\mu = p_T$. PHENIX cone criterion, including final acceptance ($-0.35 < \eta < 0.35$) was taken into account. The other calculation \cite{Vogel} uses CTEQ6M structure functions and GRV fragmentation functions and same scales $\mu = p_T$, but smaller cone radius $R=0.4$~[rad]. The smaller cone size used in \cite{Vogel} compensates neglecting of exact PHENIX acceptance so that both predictions agree with each other. Both calculations predict somewhat larger proportion of isolated photons than we find in the data being however within systematic errors of data.

\begin{figure}
\includegraphics[width=\columnwidth,height=10.5cm]{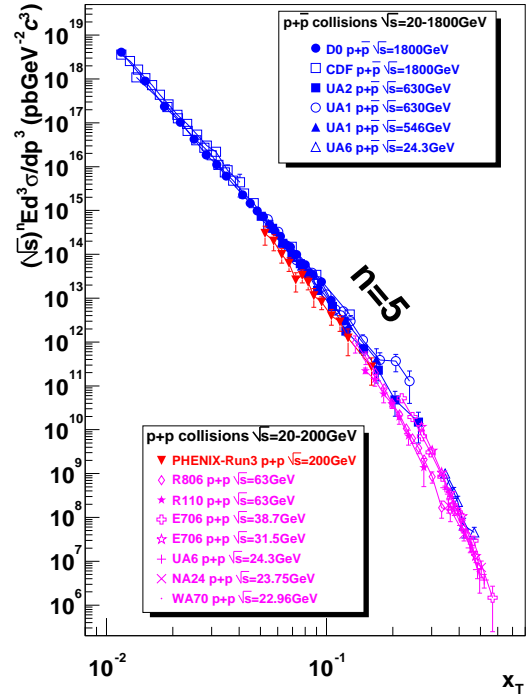} 
\caption{Comparison of world data on direct photon production in $p+p$ and $p+\bar p$ collisions scaled with $(\sqrt{s})^5$ (D0 \cite{D0}, CDF \cite{CDF}, UA1 \cite{UA1}, UA2 \cite{UA2}, UA6 \cite{UA6}, R806 \cite{R806}, R110 \cite{R110}, E706 \cite{E706}, NA24 \cite{NA24}, WA70 \cite{WA70}).}
\label{fig:xt-scale}
\end{figure}

Finally, we compare our inclusive direct photon production cross-section in p+p collisions with world data. From pQCD we know, that if there were no any additional scale in the problem, the cross-section could be expressed as $d^3\sigma/dyd^2p_T=(\sqrt{s})^4\cdot F(x_T,y)$, where $F(x_T,y)$ some dimensionless function and $x_T=2p_T/\sqrt{s}$. In the reality higher order corrections ($Q^2$ dependence of $\alpha_s$ and structure functions) result in larger power of $\sqrt{s}$. We present comparison of all available word data on photon production in $p+p$ and $p+\bar p$ collisions scaled with power $n=5$ in fig.~\ref{fig:xt-scale}. We find good agreement of our results with other world data. 

\begin{figure}
\includegraphics[width=\columnwidth]{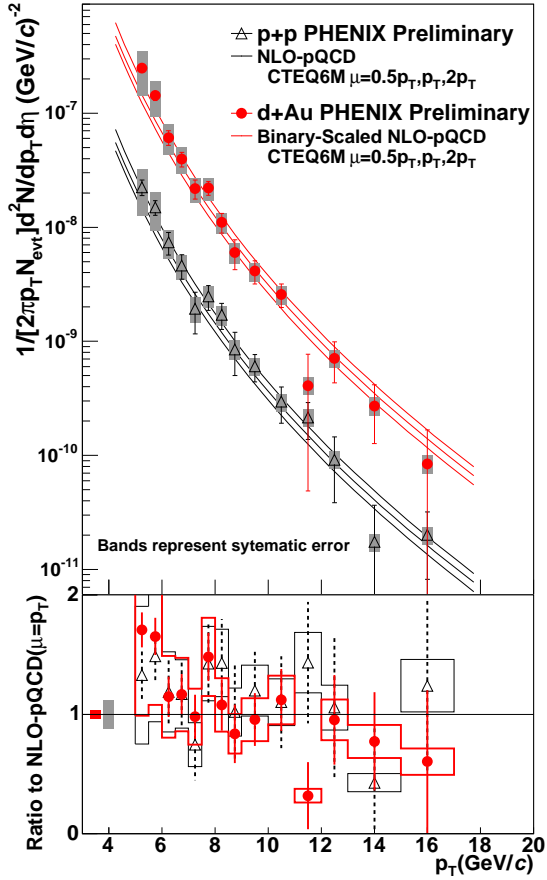} 
\caption{Inclusive direct photon yield, measured in minimal bias d+Au collisions at $\sqrt{s_{NN}}=200$ GeV compared to inclusive direct photon yield in p+p collisions and NLO pQCD predictions \cite{Vogel} scaled with number of binary collisions.}
\label{fig:dAu-spectr}
\end{figure}

We explore influence of cold nuclear matter measuring direct photon yield in d+Au collisions.  The preliminary direct photon data for minimum-bias d+Au collisions at $\sqrt{s_{NN}}=200$~GeV are presented in fig.~\ref{fig:dAu-spectr}. These data are based on $\approx 3$ billion events sampled in RHIC Run-3. For comparison we present as well direct photon yield measured in p+p collisions at the same energy and pQCD predictions \cite{Vogel} scaled with the number of binary collisions. The more detailed comparison is presented in the bottom part of the figure, where ratios of data to pQCD predictions with scales set to $\mu=p_T$ are shown for p+p and d+Au cases. In both cases we find good agreement with pQCD predictions in the full $p_T$ range.

\begin{figure}
\includegraphics[width=\columnwidth]{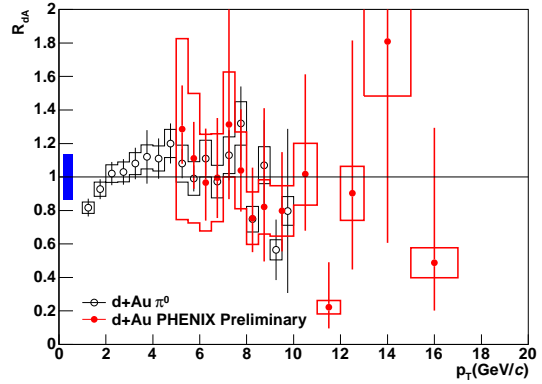} 
\caption{Nuclear modification factor of direct photons (solid circles) and neutral pions (empty circles) measured in minimal bias d+Au collisions at $\sqrt{s_{NN}}=200$~GeV.}
\label{fig:dAu-rat}
\end{figure}

To quantify modification of particle emission in d+Au collisions we use nuclear modification factor: % $R_{dAu}$: 
$$
R_{dAu}=\frac{\frac{d^3N_{dAu}}{dy\,d^2p_T}}{\frac{<N_{coll}>}{\sigma_{inel}}\frac{d^3\sigma_{pp}}{dy\,d^2p_T}}.
$$
We present comparison of nuclear modification factors calculated in minimal bias d+Au collisions for neutral pions and inclusive direct photons in fig.\ref{fig:dAu-rat}. In the case of pions we find small increase of $R_{dAu}$ at $p_T>2$~GeV, while for direct photons nuclear modification factor agrees within errors both with unit and with pion $R_{dAu}$. 

To conclude, PHENIX measured direct photon yield in p+p and d+Au collisions at $\sqrt{s_{NN}}=200$~GeV. Inclusive direct photon production cross-section measured in p+p collisions agrees with pQCD predictions in all $p_T$ range. For the first time PHENIX measured simultaneously inclusive and isolated direct photon yield in p+p collisions. Relative contribution of isolated photons is somewhat smaller than pQCD predictions but still agrees with them within errors. Comparing direct photon yields in d+Au and p+p collisions we find that data agree within errors with absence of modifications due to presence of cold nuclear matter.

\vbox{

}
\end{document}